%% file: letter_final.tex
\documentclass[preprintnumbers,amsmath,amssymb,nofootinbib,twocolumn]{revtex4-1}

\usepackage{dcolumn}
\usepackage{bm}
\usepackage{graphicx}
\usepackage{subcaption}

\usepackage{color}
\usepackage{natbib}
\usepackage{twoopt}
\input{newcommands.tex}

\usepackage[pdftex,
            breaklinks=true,%
            colorlinks=true,%
            pdfauthor={Lavalle, Maurin, Putze},%
            pdftitle={Template for article in XXX}%
           ]{hyperref}

\graphicspath{{./}{Figs/}}

\begin{document}

\title{Direct constraints on diffusion models from cosmic-ray positron data:
Excluding the Minimal model for dark matter searches}
\author{Julien Lavalle}
\affiliation{Laboratoire Univers \& Particules de Montpellier (LUPM),
  CNRS-IN2P3 \& Universit\'e Montpellier II (UMR-5299),
  Place Eug\`ene Bataillon,
  F-34095 Montpellier Cedex 05 --- France}
\email{lavalle@in2p3.fr}

\author{David Maurin}
\affiliation{LPSC, Universit\'e Grenoble-Alpes, CNRS/IN2P3,
      53 avenue des Martyrs, F-38026 Grenoble --- France}
\email{dmaurin@lpsc.in2p3.fr}

\author{Antje Putze}
\affiliation{LAPTh, Universit\'e de Savoie, CNRS, 
  9 chemin de Bellevue B.P.110,  F-74941 Annecy-le-Vieux --- France
}
\email{putze@lapth.cnrs.fr}

\begin{abstract}
Galactic Cosmic-ray (CR) transport parameters are usually constrained by the boron-to-carbon 
ratio. This procedure is generically plagued with degeneracies between the diffusion coefficient 
and the vertical extent of the 
Galactic magnetic halo. The latter is of paramount importance for indirect dark matter (DM) 
searches, because it fixes the amount of DM annihilation or decay that contributes to the 
local antimatter CR flux.
These degeneracies could be broken by using secondary radioactive species, 
but the current data still have large error bars, and this method is extremely sensitive to 
the very local interstellar medium properties.
Here, we propose to use the low-energy CR positrons in the GeV range as another direct
constraint on diffusion models. We show that 
the PAMELA data disfavor small diffusion halo ($L\lesssim 3$ kpc) and large diffusion 
slope models, and exclude the minimal ({\em min}) configuration (Maurin et al. 2001, 
Donato et al. 2004) widely used in the literature to bracket the uncertainties in the 
DM signal predictions. This is complementary to indirect constraints (diffuse radio and gamma-ray
emissions) and has strong impact on DM searches. Indeed, this makes the 
antiproton constraints more robust while enhancing the discovery/exclusion potential of 
current and future experiments, like AMS-02 and GAPS, especially in the antiproton and 
antideuteron channels.
\end{abstract}

\maketitle
\preprint{LUPM:14-019}

\section{Introduction}
The theoretical understanding of cosmic-ray (CR) transport relies on diffusion of charged particles 
off magnetic turbulences and has been established for decades 
\cite{Berezinskii1990,Schlickeiser2002,Strong2007}. In this picture, CRs are confined in an 
extended region that encompasses the Galactic disk, which can be assumed as
a homogeneous magnetic cylinder at first order. Therein the diffusion tensor is reduced to a 
rigidity-dependent scalar (homogeneous and isotropic diffusion). Yet, it is only 
very recently that we have been able to start probing the fine structure of CR 
phenomenology. With the advent of space experiments like PAMELA 
\cite{Adriani2009,Adriani2010b,Adriani2011a,Adriani2011c,Adriani2013a,Adriani2013}, 
Fermi \cite{Abdo2009c,Ackermann2012j}, and more recently 
AMS-02 \cite{Aguilar2013}, the physics of Galactic CRs has just entered 
the precision era. CR measurements also provide very interesting probes of exotic physics. 
In particular, the survey of antimatter CR species may unveil traces of dark 
matter (DM) annihilation or decay in the Galaxy ({\em e.g.}, Refs.~\cite{Silk1984,Porter2011,Lavalle2012a}).

\modif{The background to DM searches mostly comes from secondary CRs, {\em i.e.} those CRs
produced from nuclear interactions between the CR nuclei and the interstellar gas}. This secondary 
component is used to constrain the CR transport model parameters, as the ratio of 
secondary-to-primary CR nuclei depends very little on the properties of the primaries at their 
sources, while very strongly on the transport 
history \cite{Maurin2002,Putze2011}. The most widely used ratio is B/C \cite{Webber1992,Jones2001a,Maurin2001,Strong2001,Lionetto2005,Evoli2008,Putze2010,Trotta2011}, 
although other ratios like $^2$He/$^1$H and $^3$He/$^4$He are equally powerful 
\cite{Coste2012}. Once the transport parameters are set (from B/C analysis), one can 
fully predict the fluxes of the other secondary species \modif{(same propagation history)} 
provided the relevant production cross 
sections are known. Such calculations have been done for secondary positrons 
\cite{Moskalenko1998,Delahaye2009a,Delahaye2010,Lavalle2011a,DiMauro2014}, 
antiprotons \cite{Donato2001,Bringmann2007b,Donato2009,DiBernardo2010}, and 
antideuterons \cite{Donato2008}. For all but the positron case, for which energy 
losses play a major role in contrast to nuclei, these computations are poorly sensitive to the 
theoretical uncertainties affecting the transport parameters in spite of the large
degeneracies induced by the B/C analysis \cite{Jones2001a,Maurin2001}.

In 2-zone diffusion models, a large uncertainty for DM searches stems from the degeneracy 
between the normalization of the diffusion coefficient\footnote{We assume 
$K({\cal R}\equiv |p/q|) = \beta \, K_0 ({\cal R}/1\,
{\rm GV})^\delta$ in the following --- where $p$ is the momentum, $\beta$ the velocity, and $q$ the 
electric charge.} $K_0$
and half the vertical extent of the diffusion halo $L$. Indeed, the B/C data mostly constrain the 
CR escape time $\propto L/K_0$ --- as all other secondary-to-primary ratios of stable 
nuclei originating from the Galactic disk. In contrast, DM-induced CRs are produced all 
over the diffusion halo (and outside), and its size $L$ has a 
strong impact on the signal predictions: their flux roughly scales like $\sim L^2/K_0$ (assuming a 
constant DM density, a fairly good approximation for qualitative understanding). This picture is 
valid whenever the transport is dominated by spatial diffusion.

In Ref.~\cite{Donato2004}, the Authors proposed two extreme configurations to bracket 
the theoretical uncertainties on the DM signal predictions, dubbed {\em min} and 
{\em max}, relying on the B/C analysis performed in \cite{Maurin2001}. The former (latter) is 
featured by a 
very small (large) diffusion zone with $L=1$ (15) kpc, and is associated with low (large) signal 
predictions. In practice, the {\em min} model is usually invoked to minimize the antiproton 
constraints on DM candidates, while {\em max} is used to promote detectability --- the relative 
difference between the two almost reaches two orders of magnitude in terms of flux predictions. 
Such a large range for $L$ strongly affects the antimatter CRs as reliable probes of the 
DM parameter space. This is particularly important in the light-intermediate WIMP mass range 
(10-100 GeV), where antiprotons could be used to place severe constraints on WIMPs annihilating or 
decaying into quarks~\cite{Lavalle2010d}.

There are serious hints that $L$ should be larger than $\sim$1 kpc, but no stringent bounds so far. 
Radioactive species are insensitive to $L$ at low energy because their lifetime is shorter
than the diffusion time to reach the halo boundary. Using for instance $^{10}$Be/$^{9}$Be breaks
the $K_0/L$ degeneracy and sets constraints on $L$ \cite{Strong2001,Jones2001a}. However, it was 
shown in several studies that this method is very sensitive to the modeling of the local 
interstellar medium (ISM), and strongly affected by the presence of a local under-density, known 
as the {\em local bubble} \cite{Lallement2003,Lallement2014}: this relaxes the 
lower bound on $L$, depending on the size of the under-dense region \cite{Donato2002,Putze2010}. 
There are also other, while more indirect, hints for larger values for $L$, coming {\em e.g.} from
calculations of the diffuse Galactic gamma-ray \cite{Ackermann2012e} or radio emissions \cite{Orlando2013,Bringmann2012,DiBernardo2013}. Nevertheless, 
predictions of these observables rely on more ingredients (line-of-sight integrals depending
on the astrophysical source, ISM, and/or magnetic field distributions).

In this paper, we propose to use the low-energy {\em secondary} CR positrons as an additional 
direct constraint on $L$. We exploit the fact that the propagation history 
of positrons is 
different from that of nuclei, due to energy losses. This typically shortens the mean free path of 
positrons, and the dependence on $L$ is milder. In particular, the secondary positron flux roughly 
scales like $\sim 1/\sqrt{K_0}$, allowing us to place a lower bound on $L$ from the current 
positron data, assuming the B/C-induced relation between $K_0$ and $L$. We will only rely on 
secondary positrons, though it is clear that a primary 
component is also expected from recent measurements of the positron 
fraction \cite{Adriani2009,Abdo2009c,Aguilar2013}. Our approach,
suggested in \cite{Lavalle2012}, is complementary to the study carried out later in 
Ref.~\cite{DiBernardo2013}, though with a different propagation
treatment, for which the main constraints came from the diffuse radio emission data. 
We first briefly discuss the propagation modeling and relevant parameters, then sketch our
statistical analysis method before going to the results and conclusion.
\section{Transport of cosmic-ray positrons}
\label{sec:positrons}
We wish to constrain small 2D diffusion halo models, with $L \sim 1$ kpc. 
In this context, as observers located at 8 kpc from the Galactic center while far enough from 
the radial border located at $\sim$ 10-15 kpc from us, we can neglect the radial 
escape. We then restrict ourselves to a much simpler 1D problem along the vertical axis. Secondary 
positrons originate in the Galactic disk from \modif{inelastic scattering processes of 
primary CRs off the interstellar gas}. This source term is well constrained inside a radius 
$\sim 1$ kpc around the Earth since the CR flux barely varies over such a distance, and the 
average gas density confined in the disk is well estimated. It can be approximated to 
${\cal Q}(E,\vec{x}) = 2\,h\,n_{\rm ism}\,\delta(z)\,Q_0(E)$, where $n_{\rm ism} = 1 \,{\rm cm^{-3}}$
is the ISM gas density, and $h=100$ pc is half the disk width. The energy dependence is carried by 
\modif{$Q_0(E)=4\,\pi \,\sum_{i,j} f_j \int dT \,(d\phi_{\rm cr,i}(T)/dT) 
\,(d\sigma_{ij\to e^+}(E)/dE)$}, which
convolves the CR flux (species $i$) with the ISM gas (species $j$, featuring a fraction $f_j$).

In the following, we will stick to the formalism presented in \cite{Delahaye2009a,Delahaye2010,Lavalle2011a} for the calculation of secondary positrons. In \cite{Delahaye2009a} (see their Fig. 10), 
it was shown that transport configurations implying both reacceleration and convection resulted in 
a prominent low-energy bump around 1 GeV in the secondary positron flux. To be conservative in our 
comparisons with the data, we consider that positrons are only driven by spatial diffusion and 
energy losses. This actually significantly reduces the computational time. We can define an 
energy-dependent propagation scale $\lambda$ as:
\ben
\lambda^2 (E,E_s) & = & 4 \int_E^{E_s} dE'\,\frac{K(E')}{b(E')} \\
\label{eq:lambdadef}
& = &
(3.56 \,{\rm kpc})^2 \, \frac{K_0}{10^{-2}\,{\rm kpc^2/Myr}} \, 
\frac{\tau_l}{10^{16} \,{\rm s}}\,\nn\\ 
& \times & \left\{ \frac{(E/E_0)^{(\delta-\alpha+1)}}{\alpha-\delta-1} 
\left[ 1-(E/E_s)^{\alpha-\delta-1} \right] \right\}\,,
\label{eq:lambdanum}
\een
where $E$ ($E_s\geq E$) is the observed (injected) positron energy, and where we have used
$b(E) = -dE/dt = (E_0/\tau_l)(E/E_0)^\alpha $ for the energy losses. For GeV 
positrons, losses are dominated by inverse Compton and synchrotron processes and the Thomson 
approximation holds, such that $\alpha = 2$ and $\tau_l\simeq 10^{16}{\rm s}$ (for $E_0=1$ GeV)
to a very good approximation \modif{(we will use a more accurate form for $b(E)$ in practice)}. The 
propagation scale $\lambda$ allows us to write the Green function for the positron transport:
\ben
{\cal G}_{1{\rm D}}(E,z\leftarrow E_s,z_s) = \frac{e^{-|\frac{z-z_s}{\lambda}|^2}}
{b(E) \,\sqrt{\pi\,\lambda^2}} \overset{z,z_s\to 0}{\longrightarrow} \frac{1}{b(E) \,\sqrt{\pi\,\lambda^2}}\,.
\een
For small-$L$ models, boundary effects are important. When both
the observer and the source are confined into the disk ($z,z_s=0$), the following series
expansions can be used in the regime $\lambda\gtrsim L$, relevant at low energy \cite{Lavalle2007a}:
\ben
{\cal G}_{1{\rm D, Helm}}(E\leftarrow E_s) &=& \frac{1}{b(E)\,L}\sum_{n=1}^{n=+\infty}
e^{ - \left| \frac{(2\,n-1)\,\pi\,\lambda}{4\,L}\right|^2 } \\
& \approx & 
\frac{   e^{ - \left|\frac{\pi\,\lambda}{4\,L}\right|^2 }  + 
e^{ - \left|\frac{3\,\pi\,\lambda}{4\,L}\right|^2 } }{b(E)\,L}\nn\,,
\een
If we express the B/C correlation between $K_0$ and $L$ as $K_0 = \kappa\, L$,
with $\kappa\sim 10^{-3}$ kpc/Myr \cite{Maurin2001}, we can read off the dependence of the CR 
positron density on $L$ from the leading term:
\ben
{\cal G}_{1{\rm D, Helm}}(E\leftarrow E_s) 
\approx 
\frac{  e^{ -  \frac{\kappa \,\tau_l \, f(E,E_s)}{L} } }{b(E)\,L}\nn\,,
\een
where $f(E,E_s) \overset{\sim}{\propto} (E/E_0)^{(\delta-1)}/(1-\delta) \approx 2 
/ \sqrt{E/({\rm GeV})}$.
Interestingly, the pre-factor is proportional to $1/(b(E)\,L)$, where it appears 
that a small $L$ may lead to a large secondary positron flux.

The additional impact of $\delta$ can be understood from the spectral shape predicted for the 
secondary positron flux \cite{Delahaye2010}, which roughly scales as $E^{-\tilde{\gamma}}$, with the 
spectral index $\tilde\gamma \simeq \gamma+(1+\delta)/2$, where $\gamma\approx 2.7$ is the source 
index associated with secondary positron production. Therefore, the smaller $\delta$ the harder
the secondary positron spectrum.

Finally, we note that since we focus on the GeV energy range, our results are sensitive 
to solar modulation effects that we will include in the force-field approximation 
\cite{Gleeson1968}.
\section{Statistical analysis}
\label{sec:stat}
Given the flux data $\phi_{\text{data}} (E)$ and associated statistical error 
$\sigma_{\phi_{\text{data}}} (E)$, we wish to constrain {\em only} those transport models leading to
secondary positron fluxes {\em in excess} with respect to the data. For each data point at 
kinetic energy $E_i$, the number of standard deviations in a one-sided 
hypothesis test of a Gaussian variate, {\em i.e.} the $Z$-score, is calculated:
\begin{equation}
Z_i = \frac{\phi_{\text{model}} (E_i) - \phi_{\text{data}} (E_i)}{\sigma_{\phi_{\text{data}}} (E_i)}\,,
\end{equation}
where $\phi_{\text{model}} (E_i)$ is the modulated flux estimated for a given parameter set 
\{$K_0$, $L$, $\delta$\}. In a subsequent step, the individual p-value $p_i$ is estimated only for 
data points with a positive $Z_i$:
\begin{equation}
p_i = 1 - \Phi (Z_i) = \frac{1 - \text{erf} (Z_i/\sqrt{2})}{2}\,,
\end{equation}
where $\Phi (Z)$ is the cumulative distribution function of the Gaussian distribution. The 
independent $p_i$ values of a given model are eventually combined into a single test statistic 
$X$ using Fisher's method \cite{Fisher1970}:
\begin{equation}
X = -2 \sum_i^n \log{p_i}\,.
\end{equation}
$X$ follows a $\chi^2_{2n}$ with $2n$ degrees of freedom from which a p-value for the global 
hypothesis can be obtained:
\begin{equation}
p = 1 - \frac{\gamma (n, \chi^2_{2n}/2)}{\Gamma(n)},
\end{equation}
where $\gamma$ and $\Gamma$ are the lower incomplete and complete gamma functions, respectively.

In the following, we will use an exclusion criterion of $p<0.001345$, which corresponds to
exceeding the data by 3-$\sigma$ or more.
%
%
%
\section{Results and discussion}
\label{sec:results}
The propagation framework is that defined in \cite{Delahaye2009a,Delahaye2010}, including 
the energy loss parameters. As we disregard reacceleration and convection, a transport model is 
defined by $K_0$, $\delta$, and $L$, which we vary in the ranges $K_0/L\in\,$[$10^{-3}$,$10^{-2}$] 
kpc/Myr, $\delta\in\,$[0.2-0.9], consistent with the B/C constraints~\cite{Maurin2001}.
\begin{figure*}[!t]
\centering
\includegraphics[width = \textwidth]{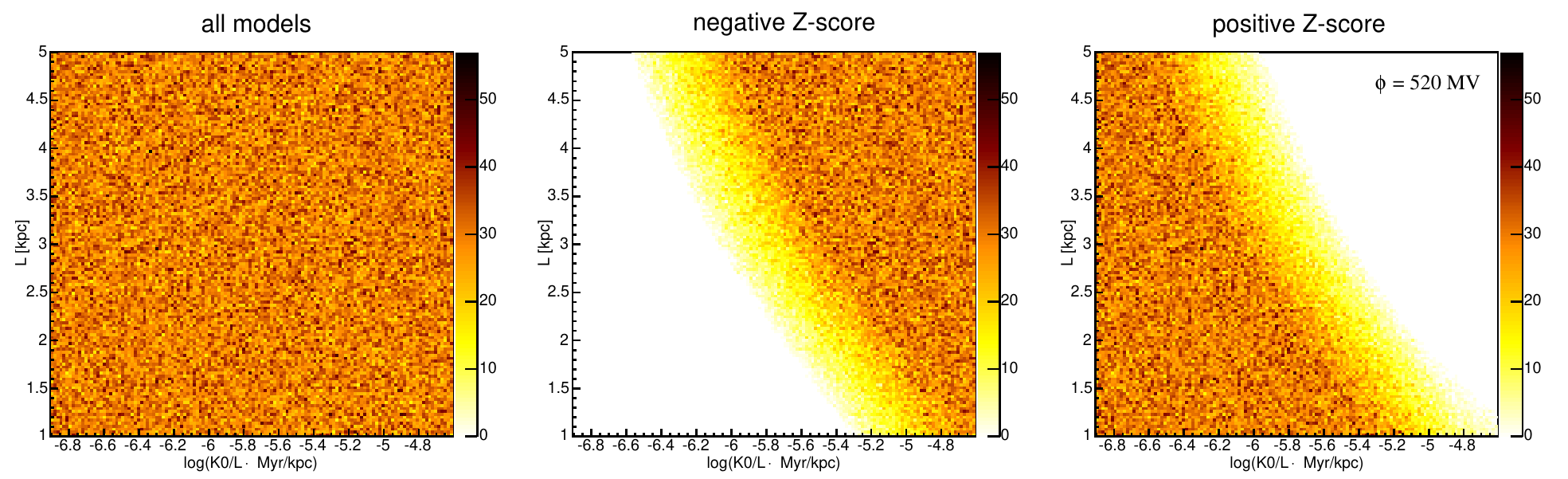}
\caption{2D histograms for $L$ vs $\log(K_0/L)$ for $\delta \in\,$[0.2-0.9]. 
  \modif{The color code indicates the density of propagation models per 2D bin.} }
\label{fig:Zscore}
\end{figure*}
\begin{figure*}[!t]
  \centering
\includegraphics[width = 0.32\textwidth]{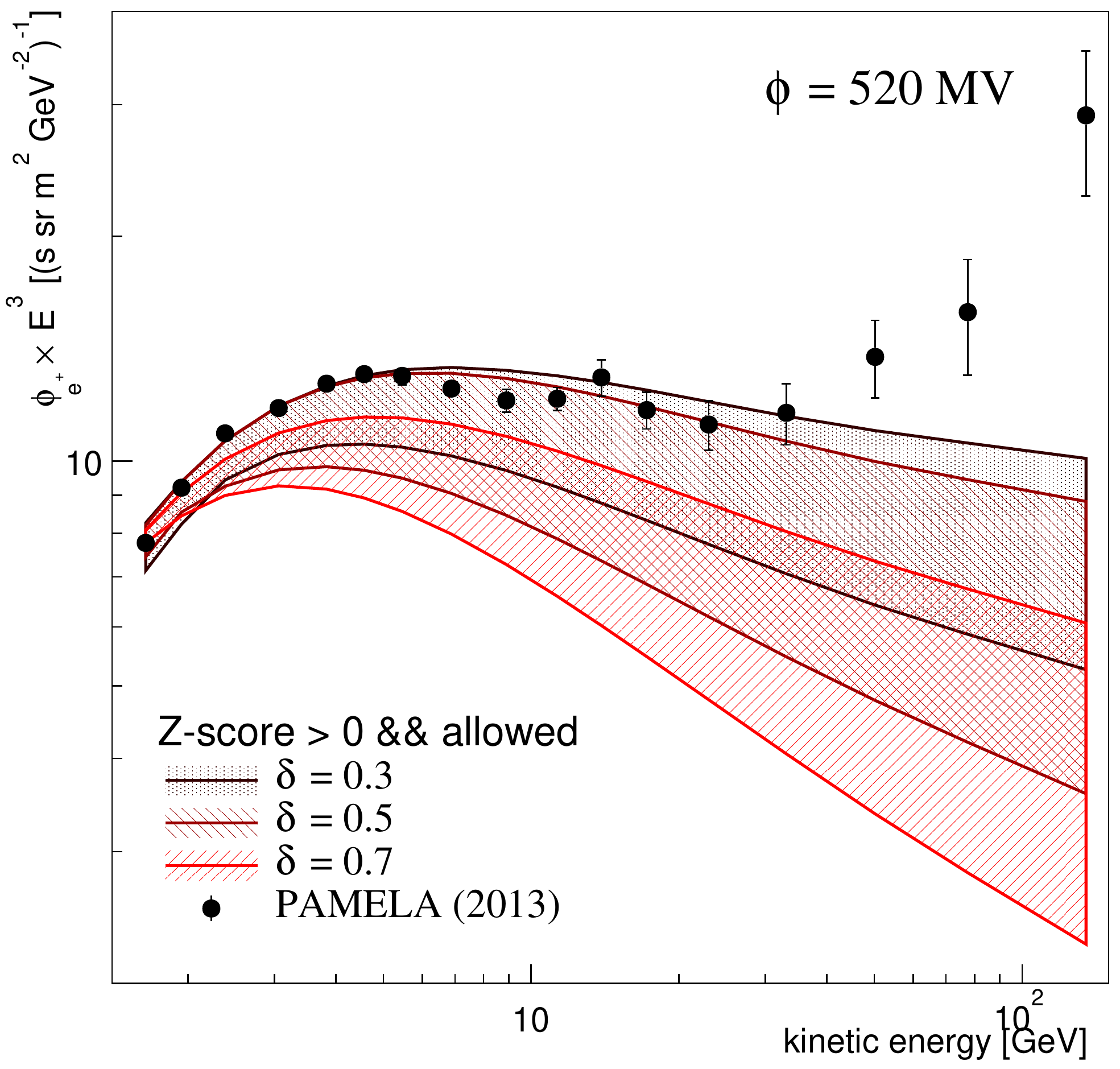}
\includegraphics[width = 0.32\textwidth]{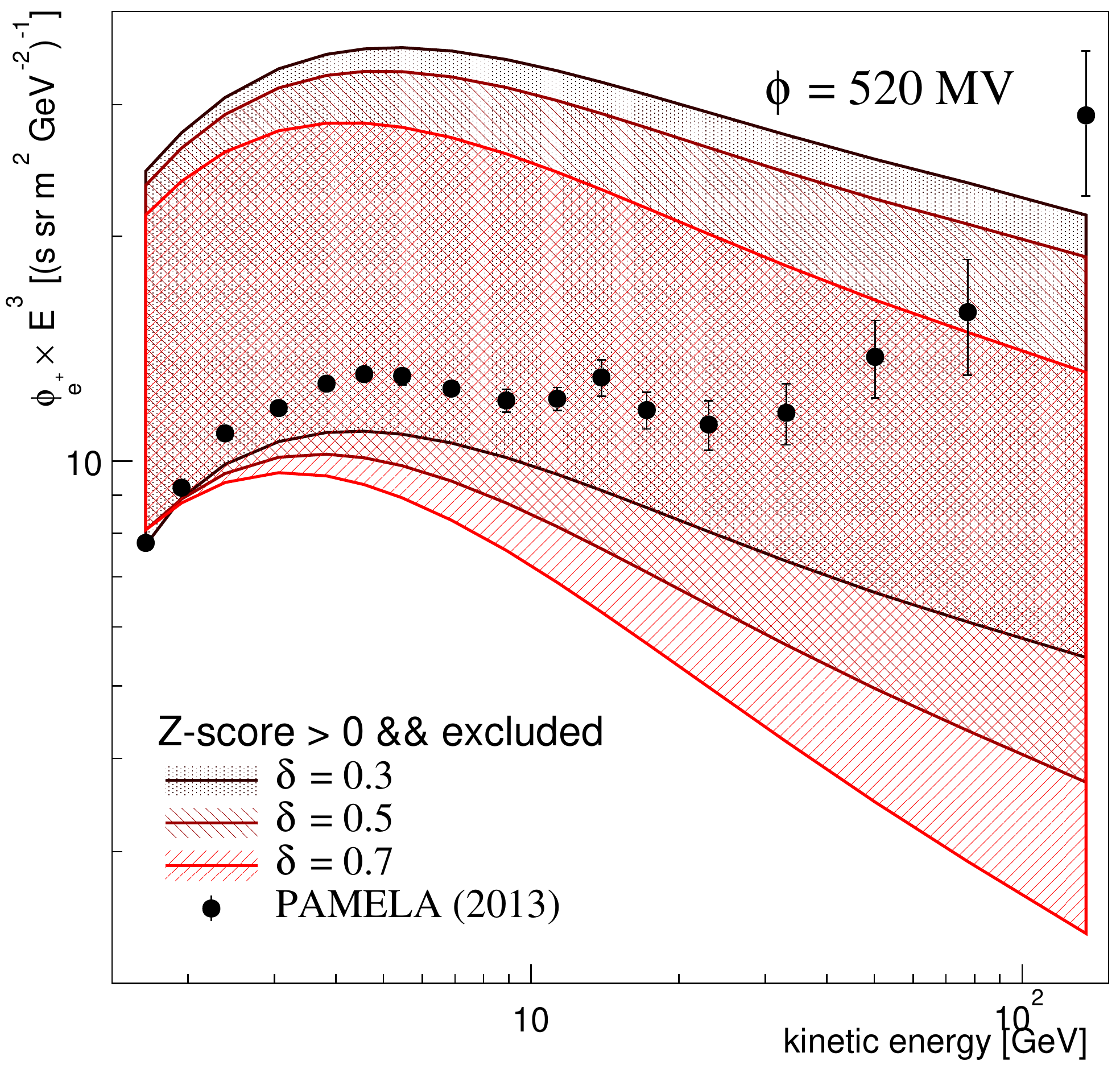}
\includegraphics[width = 0.32\textwidth]{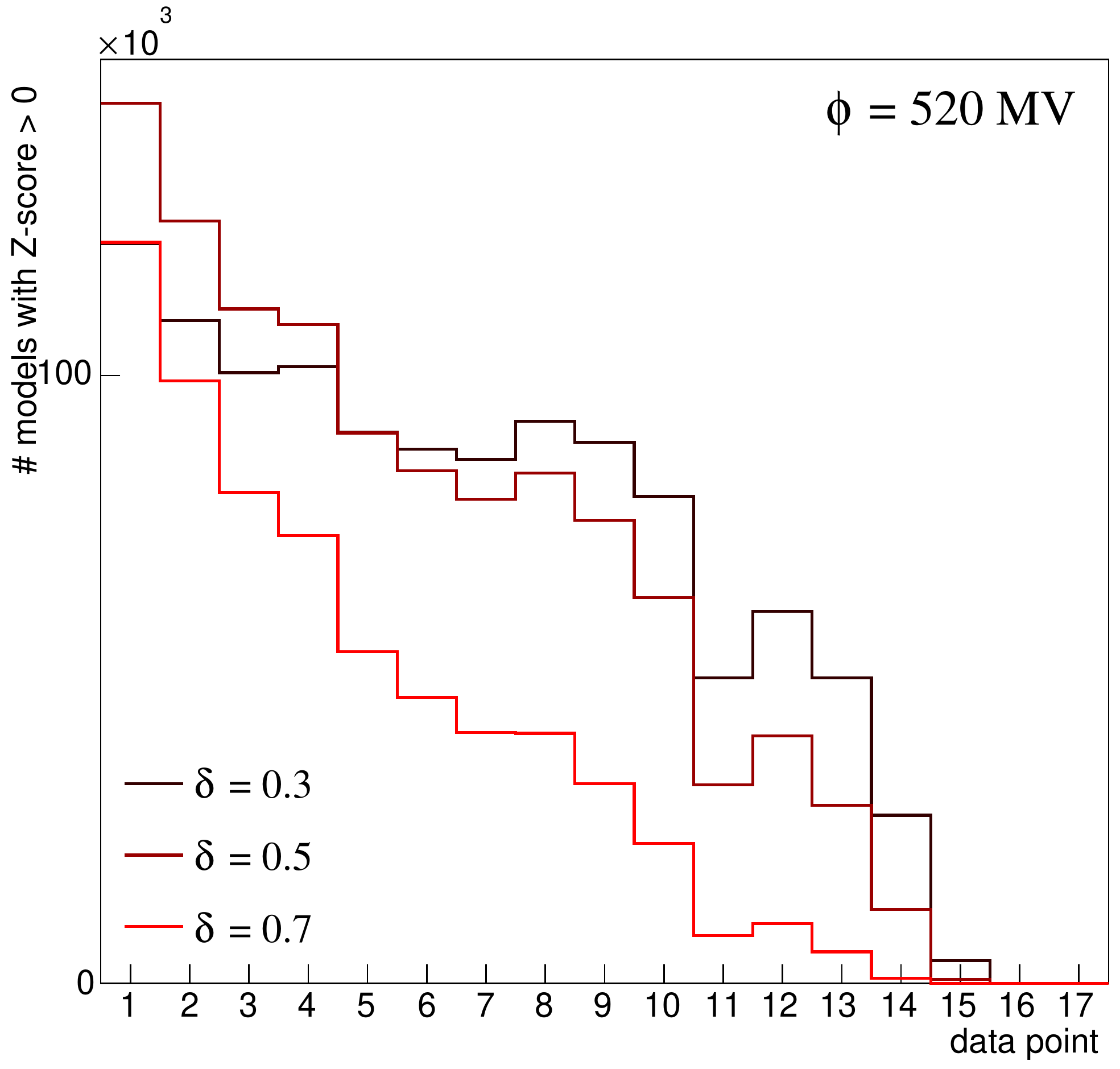}
  \caption{Left: contours of the secondary positron flux predictions in excess with one or more 
    data points, still allowed by our analysis. Middle: the same for models excluded by our 
    analysis. Right: \modif{histogram of models with $Z_i>0$ excluded by our analysis for a
      given PAMELA data point (17 data points from 1.64 to 135 GeV)}.}
  \label{fig:flux}
\end{figure*}

To cover the great variety of possible diffusion model parameters, 
500\,000 parameter sets have been uniformly drawn in the ranges defined above. We have used
the PAMELA data \cite{Adriani2013} associated with a solar modulation potential of 520 MV
relevant to the data taking period~\cite{Maurin2014}. We also checked our method with the 
unpublished AMS-02 data presented at ICRC 2013, but will only comment on the trend waiting for the 
published results. A 2D projection in the $\log{(K_0/L)} $--$L$ plane of (i) the models parameters 
drawn, (ii) those allowed by the PAMELA data ($\forall~i: Z_i \leq 0$), and  (iii) those leading to 
excess with respect to one or more data points ($\exists~i: Z_i > 0$), are shown in the left, 
middle, and right panels of \citefig{fig:Zscore}, respectively \modif{--- the color code indicates 
the density of propagation models per 2D bin}. The AMS-02 data from ICRC 2013 would lead to 
similar trends. It is clear that the positron data constrain extensively low 
values of $L$ and $\log(K_0/L)$ independent of the spectral index $\delta$ (white area in the 
middle plot).

The influence of $\delta$ can be understood from \citefig{fig:flux}. In the left (middle) 
panel, we show the predicted fluxes giving $Z>0$ while not excluded (excluded) by our 
analysis, for different slices in $\delta$ (the $Z<0$ models would give fluxes that 
spread over a large area below the data points). We see that the larger $\delta$, the more 
constraining the low-energy data points, as it can be expected from the spectral dependence
in $\delta$ (see \citesec{sec:positrons}). In contrast, predictions from a more gradual slope 
of $\delta \sim 0.3$ follow the data more closely (in the GeV range) and are hence 
\modif{more} equally constrained by the PAMELA data points 1 to 14 \modif{(\ie\ 1.64 to 33.1\,GeV,
out of a total of 17 data points up to 135 GeV)}.
In the right panel, we display the data points with $Z_i>0$ for the models excluded from our 
nominal analysis, for different slices in $\delta$. The trend explained above is explicit in 
this plot.
\begin{figure*}[!t]
\centering
\setlength{\unitlength}{\textwidth}
  \begin{picture}(1,0.3)
\put(0,0){
  \includegraphics[width = \columnwidth]{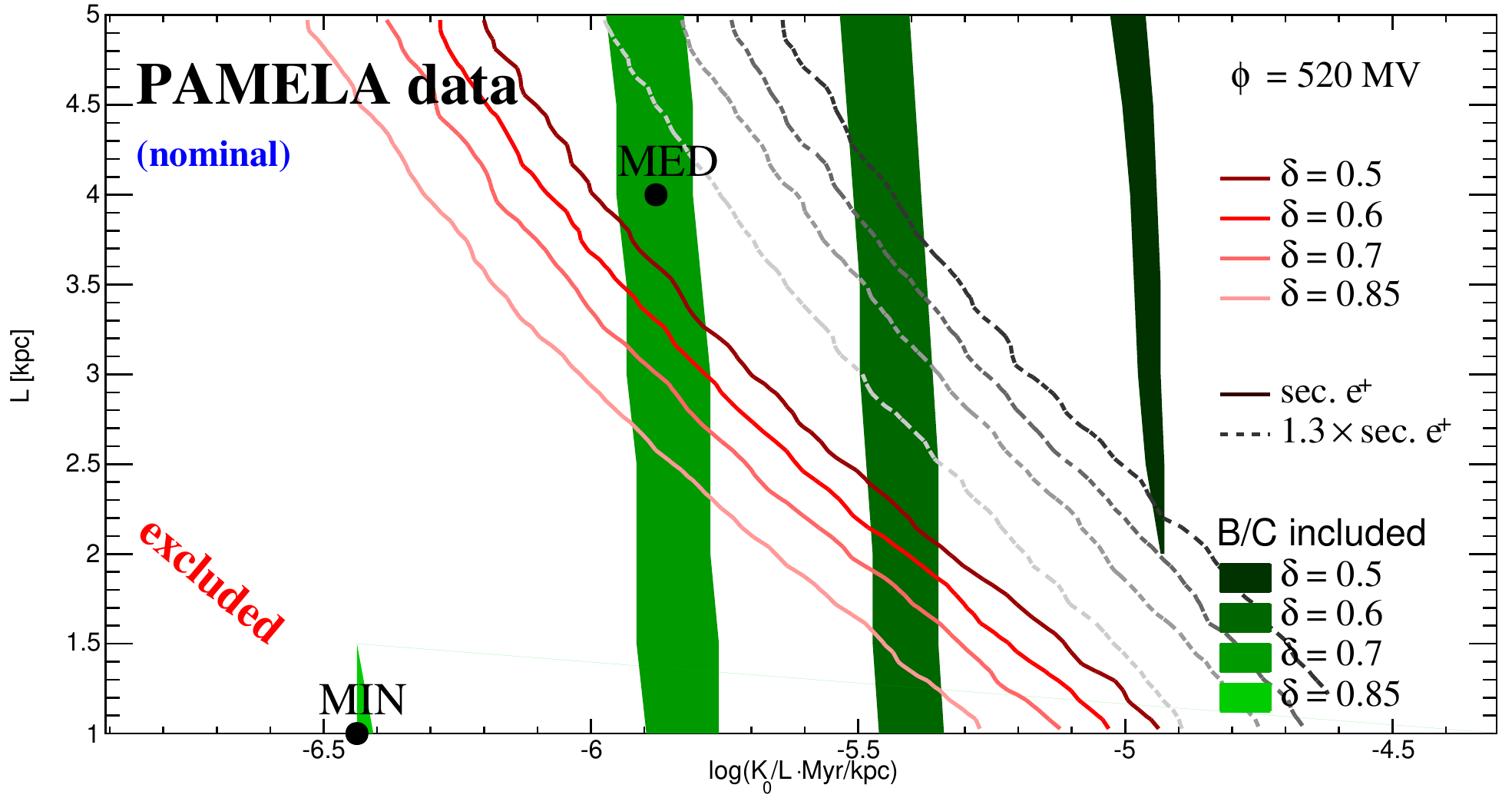}
}
\put(0.5,0){
  \includegraphics[width = \columnwidth]{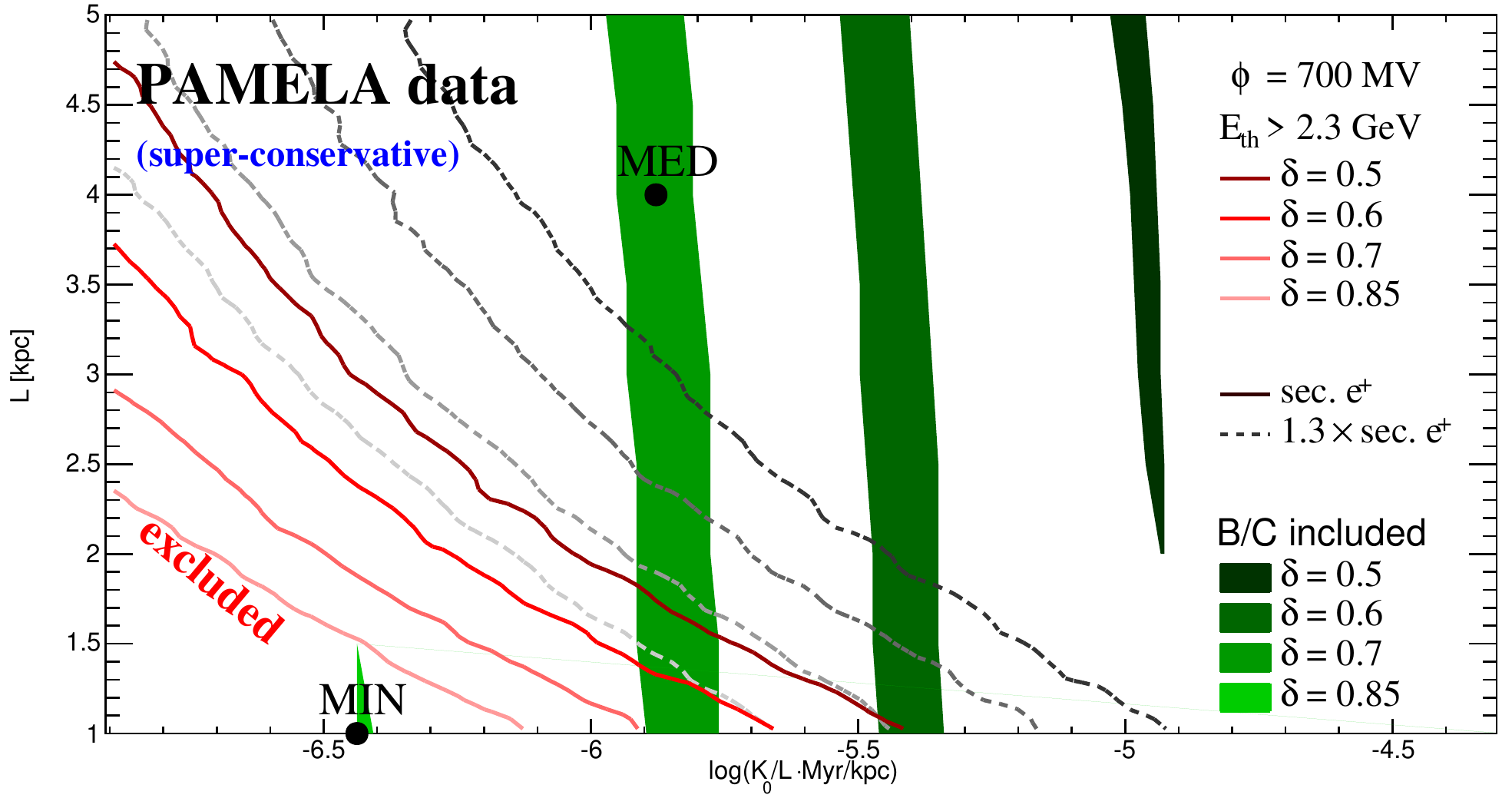}
}
  \end{picture}
  \caption{\small Constraints on propagation parameters in the $\log(K_0/L \;{\rm in}\; [{\rm kpc/Myr}])$-$ L \,[{\rm kpc}]$ plane. Lines are 
    constraints from the positron flux (downward regions are excluded). Green filled 
    contours are allowed by B/C data. The $min$ and $med$ models of 
    Ref.~\cite{Donato2004} (see Tab.~\ref{table:prop}) are indicated by a filled black 
    circle. {\bf Left panel:} positron contours (from PAMELA data) for a realistic modulation 
    level of $\phi=520$ MV. Dashed lines correspond to the limits if the secondary positron 
    prediction is increased by 30\% to mimic a primary component. 
    {\bf \modif{Right} panel (super-conservative):} Same but with a solar modulation of 700 MV 
    and considering only data points $\gtrsim$2 GeV.}
  \label{fig:exclusion}
\end{figure*}
\begin{table}
  \begin{tabular}{ccccccc} \hline
    &~~~&$\delta$&    $K_0$    &   L   & $V_c$ & $V_A$ \\
    &~~~&    -   &(kpc$^2$~Myr$^{-1}$)& (kpc) & (km~s$^{-1}$)& (km~s$^{-1}$)\\ \hline 
    max &~~~&  0.46  &   0.0765    &  15   &   5   & 117.6 \\
    med &~~~&  0.70  &   0.0112    &   4   &  12   &  52.9 \\
    min &~~~&  0.85  &   0.0016    &   1   &  13.5 &  22.4 \\
\hline
  \end{tabular} 
  \caption{Transport parameters associated with the {\em max},
    {\em med} and {\em min} DM-induced $\bar{p}$ and $\bar{d}$ fluxes.}
  \label{table:prop}
\end{table}

In \citefig{fig:exclusion}, we show the 3-$\sigma$ exclusion curves obtained in the 
$\log(K_0/L)-L$ plane, \ie\ our primary result. Plain lines are exclusion curves for different 
values of $\delta$. Dashed lines show how the result changes if the secondary flux predictions 
are increased by a global factor of 30\%, a rough way to account for a possible primary 
positron component at low energy. In the left panel, we adopt the nominal value of 520 MV 
for the solar modulation. In the right panel, we take a very conservative viewpoint to secure the 
results against systematic effects possibly coming from the solar modulation modeling or other 
low-energy effects, (i) by considering only the data above $\sim$2 GeV, and (ii) by overstating 
the solar modulation potential up to 700 MV.

We report on the same plots the B/C constraints obtained in Ref.~\cite{Maurin2001}, in the form of 
bands corresponding to different slices in $\delta$. There are several reasons as for why most 
recent studies are not used. First, although powerful statistical tools have been employed since 
\cite{Putze2010,Trotta2011,Coste2012} to sample the most probable regions of the parameter space, 
we take a wider and more conservative range such as that given in Ref.~\cite{Maurin2001}. This 
allows us to overcome possible systematic uncertainties arising in the determination of the 
transport parameters \cite{Putze2010}. Second, benchmark models $min$, $med$, and $max$ of 
Ref.~\cite{Donato2004} (see \citetab{table:prop}), widely used in the DM literature, are based on 
the parameters found in~\cite{Maurin2001}. The {\em min} and {\em med} models are shown in both
panels of \citefig{fig:exclusion}, while the {\em max} model falls outside with $L=15$ kpc. They 
belong to different B/C bands.

An important feature of \citefig{fig:exclusion} is that the positron and B/C constraints are 
almost orthogonal in the $\log(K_0/L)$-$L$ plane, and thereby uncorrelated. This is 
a strength of this approach and it can complement the one relying on radioactive species, which has 
completely different systematics. As we can see in the left panel (nominal result), the {\em min} 
model is completely excluded. This is important because this model is very often used to 
bracket the propagation uncertainties in indirect DM searches. More generally, models 
with large values of $\delta$ ($\sim 0.8$), generically associated with small diffusion halos 
($L\sim 1$ kpc), are excluded by our analysis. Smaller values of $\delta$ are featured by larger 
$K_0/L$ ratios. For example, we see that for the B/C band containing the {\em med} model 
($\delta\sim 0.7$), the positron constraint reads $L\gtrsim 3$ kpc, while a less severe 
$L\gtrsim 2.3$ kpc is obtained for smaller diffusion slopes. The dashed lines indicate how these 
limits would move if a primary positron component contributed
an additional 30\% to the secondary flux at low energy\footnote{\modif{This 30\% is the approximate 
increase needed to saturate the positron data in propagation models with $L\sim 4$ kpc, 
{\em i.e.} similar to the {\em med} configuration; spectral effects are neglected as the 
energy range relevant to get the limit is narrow.}}, leading to much larger lower bounds in $L$.
Nevertheless, though primary positrons are present in the $\sim$10-100 GeV data, their origin 
remains unclear, and their contribution at low energy can hardly be predicted.
Still, confronting the secondary positron predictions to the data provides {\em direct} 
constraints on $L$ that rely on fewer assumptions ({\em e.g.} on the Galactic magnetic or 
interstellar radiation fields) than, for example, those set from the diffuse radio or gamma-ray 
emissions, involving mostly primary electrons and positrons.

We have followed a conservative line by neglecting diffusive reacceleration, which would 
increase our predictions around 1.5-2 GeV as $\delta$ decreases \cite{Delahaye2009a}; 
we have also neglected any primary component. As we deal with low-energy positrons, the solar 
modulation modeling and intensity may also have significant impact on the limit derived. Although 
rather simplistic, the force-field approximation is expected to be accurate within $\sim$10-20\% 
above 100 MeV \cite{Potgieter2013}. In our nominal analysis, we took a modulation potential of 520 
MV, and used all PAMELA data points down to 1.64 GeV --- the lowest energy data points strongly 
constrain large-$\delta$ models. To check the robustness of our results, we adopt the radical option
(i) to increase the modulation potential up to 700 MeV and (ii) to remove the first data points
from the analysis, setting a threshold at $\sim$2 GeV (more precisely 2.38 GeV, 3$^{\rm rd}$ PAMELA 
data point). Such an option is extreme, so the corresponding results are to be taken as 
super-conservative. They are displayed in the right panel of \citefig{fig:exclusion}. The limits 
on $L$ are obviously weaker, but the {\em min} model remains excluded, as propagation models with 
large values of $\delta$ ($\sim 0.8$).

\modif{Similar limits on $L$ were obtained in other
studies, \eg\ Refs.~\cite{Trotta2011} and \cite{DiBernardo2013}. In the former,
the Authors mostly use the B/C and $^{10}$Be/$^{9}$Be ratios to constrain the CR transport 
parameters, and obtain the bound $L>3.2$ kpc (95\% CL) from a Bayesian analysis. While consistent 
with our result, this work, relying on different observables, is based on several assumptions 
that might induce biases: a low-energy spectral break in the primary source spectra and the 
absence of convection. Moreover, the presence of the local gas underdensity (local bubble) 
affects particularly this kind of studies \cite{Donato2002,Putze2010}. Our result has 
therefore his own strength as it is affected by different systematics, and may be viewed as 
slightly more conservative in the propagation setting. In Ref.~\cite{DiBernardo2013}, the Authors 
focus on the electron-positron CRs (primaries and secondaries), and derive a qualitative bound of 
$L\gtrsim 2$ kpc from 16 diffusion models (4 sets of assumptions, each taken with $L=$1, 2, 4, 
and 8 kpc). Although they briefly mention the secondary positron effect, their result relies on the 
analysis of the latitude profile of the 408 MHz radio emission, for which they further have to 
assume a model for the Galactic magnetic field. Again, our result is complementary: our propagation 
setup is more general, with different systematics, and our limit has an explicit statistical 
meaning.}
\section{Conclusion}
\label{sec:concl}
In this paper, we have investigated the low-energy positron constraints ($\lesssim$ 10 GeV) on 
2-zone CR propagation models. The obtained bounds are almost orthogonal to the B/C constraints in 
the $L$-$\log(K_0/L)$ plane, which makes them particularly attractive. The primary result is that 
we exclude the {\em min} benchmark propagation model and more generally large diffusion indices 
($\delta\gtrsim 0.8$). We also strongly disfavor small diffusion halo models with $L\lesssim 3$ 
kpc, the constraint weakening as the diffusion slope $\delta$ decreases. This has important 
consequences for DM studies that are often addressed in the frame of 2-zone models: this 
pushes the DM signal predictions toward larger values, which has a significant impact on the 
discovery/exclusion potentials of current and future experiments. This will be of particular 
interest for the searches in the antiproton and antideuteron channels with AMS-02 and GAPS 
\cite{Hailey2013,Aramaki2014,vonDoetinchem2014}.

The strength of the proposed analysis, complementary to the B/C or radioactive studies, will 
significantly improve when the AMS-02 data are released (more data points with smaller error bars). 
Preliminary calculations based on the preliminary AMS-02 data presented at the ICRC 2013 
conference already give slightly stronger constraints on $L$. The next step of this work will be 
to implement a full study including all low-energy effects and combining the coming AMS-02 data 
on positrons and PAMELA and AMS-02 data on B/C.

We note that the PAMELA collaboration has just released its B/C data \cite{Adriani2014}. With an 
analysis based on a very limited number of parameters, the Authors constrain the diffusion slope 
$\delta$ to be in the range $\sim[0.38-0.42]$. However, the choice of other free parameters (wind 
or no wind, low-energy dependence of the diffusion coefficient) has a strong impact on the result, 
and in particular on $\delta$ \cite{Putze2010,Maurin2010}.
It is important to perform a full scan of the parameter space with these new B/C data 
(awaiting the AMS-02 positron data and final B/C analysis). Better constraints on $\delta$ 
should increase the discrimination power of positrons. 

\acknowledgements{
We would like to thank L. Derome for useful discussions about statistical analysis methods.
This work was partly funded by the French ANR, {\it Programme BLANC} DMAstro-LHC,
Project ANR-12-BS05-0006, and by the {\it Investissements d'avenir, Labex ENIGMASS}.
}

\bibliography{biblio_jabref}

\end{document}

%% file: newcommands.tex
\newcommand{\modif}[1]{#1}

\newcommand{\eg}{{\it e.g.}}
\newcommand{\ie}{{\it i.e.}}

\newcommand{\citesec}[1]{Sec.~\ref{#1}}

\newcommand{\citetab}[1]{Table~\ref{#1}}
\newcommand{\citefig}[1]{Fig.~\ref{#1}}

\newcommand{\ben}{\begin{eqnarray}}
\newcommand{\een}{\end{eqnarray}}
\newcommand{\be}{\begin{equation}}
\newcommand{\ee}{\end{equation}}
\newcommand{\nn}{\nonumber}

\newcommand{\aap}{Astron.Astroph.}

\newcommand{\araa}{Ann.Rev.Astron.Astrophys.}
\newcommand{\jcap}{JCAP}